%% file: main.tex
\documentclass{article}
\usepackage{spconf,amsmath,graphicx}
\usepackage{adjustbox}
\usepackage{comment}
\usepackage{caption}
\usepackage{subcaption}
\usepackage{array}
\usepackage[colorlinks=true, pagebackref]{hyperref}
\usepackage{booktabs} 
\usepackage{tabularx}
\usepackage{enumitem}
\usepackage{cite}
\usepackage{algorithm2e}
\RestyleAlgo{ruled}
\usepackage{multirow}
\usepackage{algpseudocode}

\DeclareMathOperator*{\argmax}{arg\,max}
\DeclareMathOperator*{\argmin}{arg\,min}

\title{Patient Aware Active Learning for Fine-Grained OCT Classification}

\name{Yash-yee Logan \qquad Ryan Benkert \qquad Ahmad Mustafa \qquad Gukyeong Kwon \quad Ghassan AlRegib\thanks{This material is based upon work supported by the National Science Foundation Graduate Research Fellowship under Grant No. DGE-1650044.}}
\address{
OLIVES at the Center for Signal and Information Processing (CSIP),\\
School of Electrical and Computer Engineering,\\ Georgia Institute of Technology, Atlanta, GA, 30332-0250 USA\\ \{ylogan3, rbenkert3, amustafa9, alregib\}@gatech.edu, gukyeong.kwon@gmail.com}

\begin{document}

\onecolumn 

\begin{description}[labelindent=-1cm,leftmargin=1cm,style=multiline]

\item[\textbf{Citation}]{Y. Logan, R. Benkert, A. Mustafa, G. Kwon and G. AlRegib, "Patient Aware Active Learning for Fine-Grained OCT Classification," IEEE International Conference on Image Processing (ICIP), Oct. 2022.} \\


\item[\textbf{Review}]{Date of accept: 20 Jan 2022} \\

\item[\textbf{Codes}]{\url{https://github.com/olivesgatech/Patient-Aware-Active-Learning}} \\

\item[\textbf{Bib}] {@ARTICLE\{Logan2022\_ICIP,\\ 
author=\{Y. Logan, R. Benkert, A. Mustafa, G. Kwon and G. AlRegib\},\\ 
journal=\{IEEE International Conference on Image Processing\},\\ 
title=\{Patient Aware Active Learning for Fine-Grained OCT Classification\}, \\ 
year=\{2022\}\\ 
} \\


\item[\textbf{Copyright}]{\textcopyright 2022 IEEE. Personal use of this material is permitted. Permission from IEEE must be obtained for all other uses, in any current or future media, including reprinting/republishing this material for advertising or promotional purposes,
creating new collective works, for resale or redistribution to servers or lists, or reuse of any copyrighted component
of this work in other works. }
\\
\item[\textbf{Contact}]{\href{mailto:ylogan3@gatech.edu}{ylogan3@gatech.edu}  OR \href{mailto:alregib@gatech.edu}{alregib@gatech.edu}\\ \url{http://ghassanalregib.info/} \\ }
\end{description}

\thispagestyle{empty}
\newpage
\clearpage
\setcounter{page}{1}

\twocolumn

\maketitle

\begin{abstract}
\input{abstract}

\end{abstract}


\section{Introduction}
\input{introduction}

\section{Related Work}
\label{sec:related}

\input{relatedWork}

\section{Method}
\label{sec:multiRep}
\input{PatientAware}

\section{Experiments}
\label{sec:exp}
\input{experiments}

\section{Results}
\label{sec:results}
\input{results}

\section{Conclusion}
\label{sec:conclude}
\input{conclusion}
\label{sec:majhead}
 \clearpage
\bibliographystyle{IEEEbib}
\bibliography{refs}

\end{document}

%% file: abstract.tex
This paper considers making active learning more sensible from a medical perspective. In practice, a disease manifests itself in different forms across patient cohorts. Existing frameworks have primarily used mathematical constructs to engineer uncertainty or diversity-based methods for selecting the most informative samples. However, such algorithms do not present themselves naturally as usable by the medical community and healthcare providers. Thus, their deployment in clinical settings is very limited, if any. For this purpose, we propose a framework that incorporates clinical insights into the sample selection process of active learning that can be incorporated with existing algorithms. Our medically interpretable active learning framework captures diverse disease manifestations from patients to improve generalization performance of OCT classification. After comprehensive experiments, we report that incorporating patient insights within the active learning framework yields performance that matches or surpasses five commonly used paradigms on two architectures with a dataset having imbalanced patient distributions. Also, the framework integrates within existing medical practices and thus can be used by healthcare providers.

%% file: introduction.tex
Active learning is a branch of human-in-the-loop computing that assumes the model and unlabeled dataset will evolve over time as the most informative samples are selected from a dataset. This application applied to medical image analysis is practical as it reduces the costly and cumbersome human effort needed for expert manual annotation. Even though, active learning on medical imagery has demonstrated progress in overcoming the reliance on large, balanced datasets \cite{homeyer2011comparison, smailagic2018medal}, the practical deployment of this paradigm still has major challenges. 

\begin{figure}[tbp]
\small
\centering
\includegraphics[width =\linewidth]{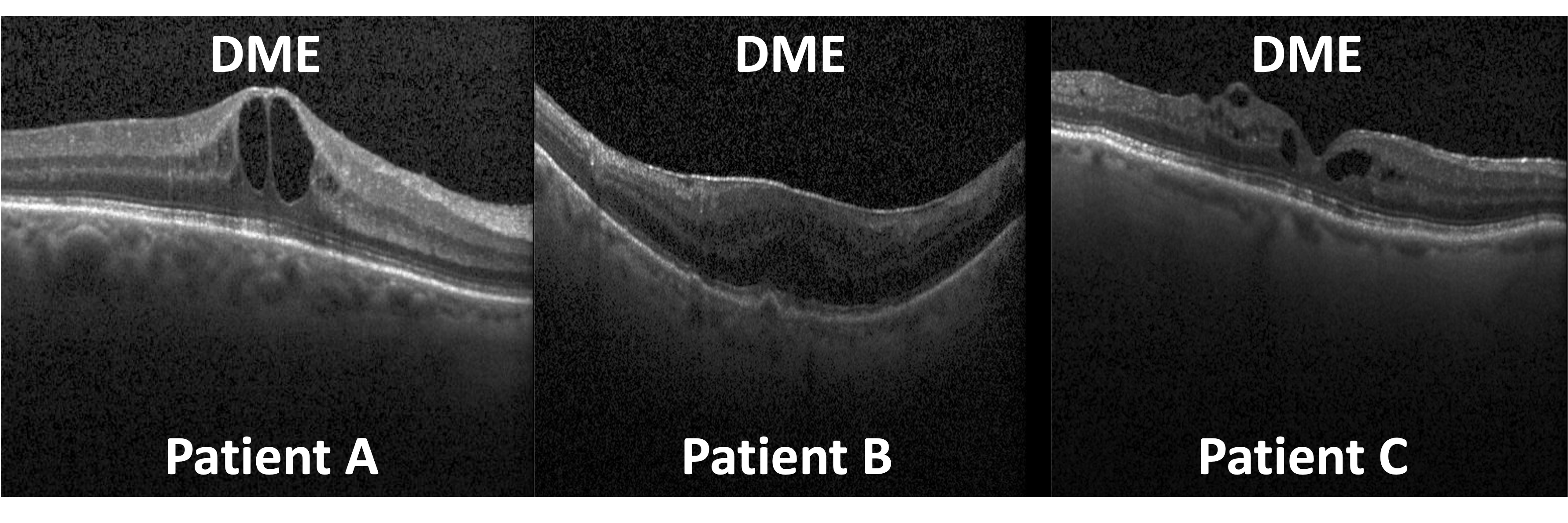}
\caption{Sample cross-sectional imagery showing differing visual characteristics of diabetic macular edema (DME) across patient cohorts.}
\label{fig:dme_pics}
\end{figure}

First, most conventional active learning paradigms applied to medical image analysis are still largely considered as ``black box" algorithms. This inhibits the health care professionals from interpreting, understanding and even correcting predictions a model has made \cite{budd2021survey, prabhushankar2021extracting, logan_multi-modal_2022}. From a physician's point-of-view, this creates a higher risk of patient harm. Second, a single disease can present itself in visually diverse formats across multiple patients. This is exemplified in Fig.~\ref{fig:dme_pics} where, although each patient's optical coherence tomography (OCT) is diagnosed with diabetic macular edema (DME), the visual characteristics of them are unalike. In other words, data diversity is captured within the medical meta data and remains un-exploited in existing active learning paradigms. Additionally, medical datasets are widely imbalanced both across classes and patients \cite{kermany2018large}. This ultimately results in existing methods training models without properly accounting for one or more patient disease manifestations from whom there is less data. For this purpose, the model could overfit to a minority patient group and fail to generalize to the broader masses. For practical deployments, this represents a major risk. 
 

\begin{figure*}[h!]
\small
\centering
\includegraphics[width =0.7\linewidth]{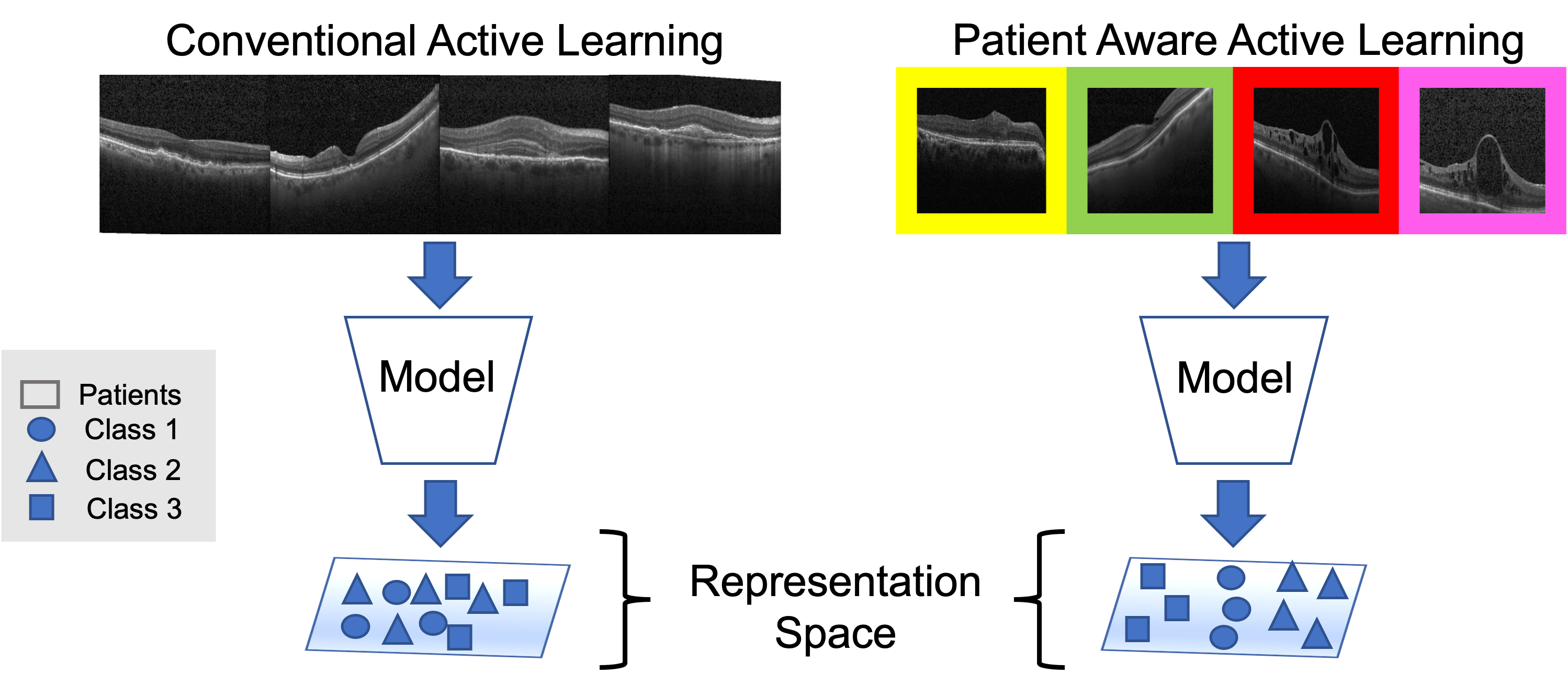}
\caption{A comparison of representation spaces between conventional and patient aware active learning paradigms.}
\label{fig:reps}
\end{figure*}

    From a machine learning perspective, the root cause of the problem resides in the structure of the representation space. Conventional active learning methods rely on a well established representation of the data. If samples are conventionally chosen from an unbalanced dataset, the structure of the representation space will be underdeveloped and the model will generalize poorly. For this purpose, we propose to shape the representation space through medical meta-information. Specifically, we provide data diversity through patient identity. By imposing a patient constraint on the query strategies, we structure the representation space based on a medically grounded prior (Fig. \ref{fig:reps}). We show that our method is robust to both unbalanced data distributions and architectures and is also effective at fine-grained disease classification. Based on the original paradigm, we define our method as \emph{Patient Aware Active Learning}. In summary, the contributions of this paper are as follows:
\begin{enumerate}[label=\roman*, leftmargin=0.4cm]
    \item We incorporate patient information to augment the active learning paradigm and improve medical interpretability.
    \item We develop a modular plug-in method that can be applied to arbitrary active learning strategies that match or outperform existing methods in terms of generalization.
    \item We test our method on a popular OCT benchmark with two different architectures and five different query strategies. This amounts to 100 experiments.
\end{enumerate}

\begin{figure*}[t]
\small
\centering
\includegraphics[width =0.85\linewidth]{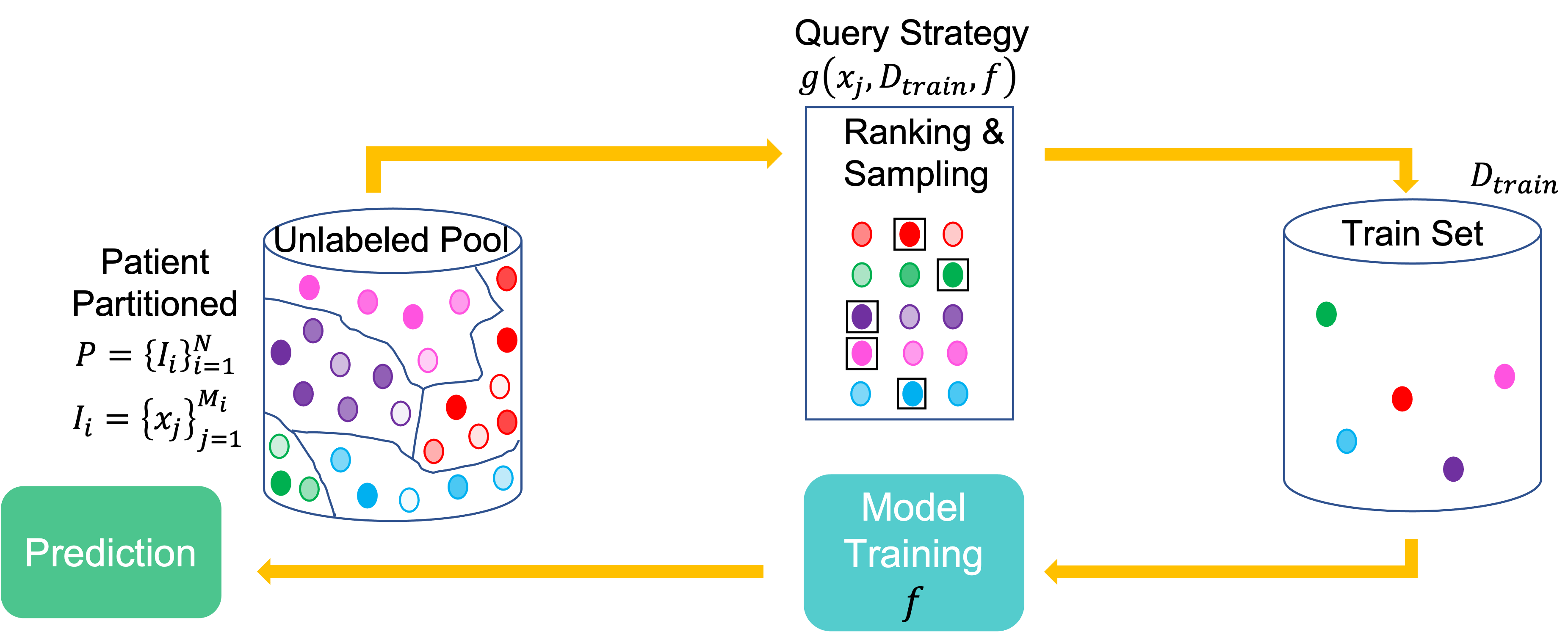}
\caption{Workflow of the patient aware active learning framework.}
\label{fig: block_diagram}
\end{figure*}



%% file: relatedWork.tex
Active learning frameworks strive to select the most informative subset of samples from an unlabeled dataset in order to combat the laborious, costly and time-consuming nature of developing large annotated datasets \cite{wang2016cost, ducoffe2018adversarial, sener2017active, gissin2019discriminative, mustafa2021man}. In relation to medical image analysis, active learning has been applied to histopathological image analysis \cite{homeyer2011comparison}, skin lesion segmentation \cite{gorriz2017cost}, heart magnetic resonance imaging and computerized tomography scan analysis \cite{pace2015interactive}, the diagnosis of digital mammograms \cite{zhao2018minimization}, and for tuberculosis detection of chest radiography \cite{melendez2015combining}. A significant area of active learning research is focused on defining sample informativeness. In this context, several approaches define informativeness with generalization difficulty \cite{zhu2009active, settles2009active} while others focus on data diversity within the acquisition batch \cite{sener2017active, gissin2019discriminative}. Even though diversity is considered within the context of learned model features, existing approaches do not account for diversity inherent within the medical data. To the best of our knowledge, our approach represents the first to consider medical meta-information within the active learning workflow.

%% file: PatientAware.tex


In a modular context, our  method  integrates within existing query  strategies  and  enhances  their  performance with medical meta-information. Patient awareness is injected into the learning framework by partitioning the unlabeled training pool on patient identity prior to ranking sample informativeness with a query function. It is a modular method that integrates with arbitrary active learning strategies.

\subsection{Patient Aware Active Learning}
For patient aware active learning (Fig.~\ref{fig: block_diagram}), we partition the unlabeled pool into the separate patients and sample from the respective subset with an arbitrary query strategy. Mathematically, we partition the unlabeled pool into the patients

\begin{align}
    \mathcal{P} = \{ I_{i}\}^{N}_{i=1}
\end{align}

\noindent where $I_{i}$ refers to the set of $M_{i}$ image-label pairs $\{(x_{j}, y_{j})\}$ and $N$ to the total amount of patients in the dataset. We begin the process by training a model $f$ on randomly selected samples from the unlabeled pool. In each following round, we choose $K$ unique patients from the set $\mathcal{P}$ (e.g., $\{I_{k}\}^{K}_{k=1}$) and sample a single image-label pair from each of the $K$ patients using the query strategy $g(x_j, D_{train}, f)$. Finally, we append the selected samples to the training pool $D_{train}$. This process is repeated to determine the minimum number of labeled samples that maximizes the model's performance.

%% file: experiments.tex
\begin{figure}[h!]
\small
\centering
\includegraphics[width =\linewidth]{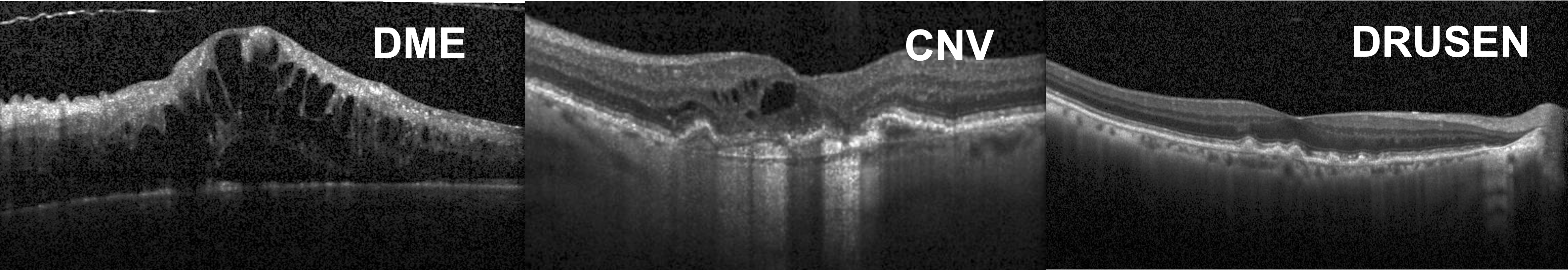}
\caption{Cross-sectional OCT of the three retinal diseases.}
\label{fig:oct_pics}
\end{figure}
\begin{figure}[h!]
\small
\centering
\includegraphics[width =\linewidth]{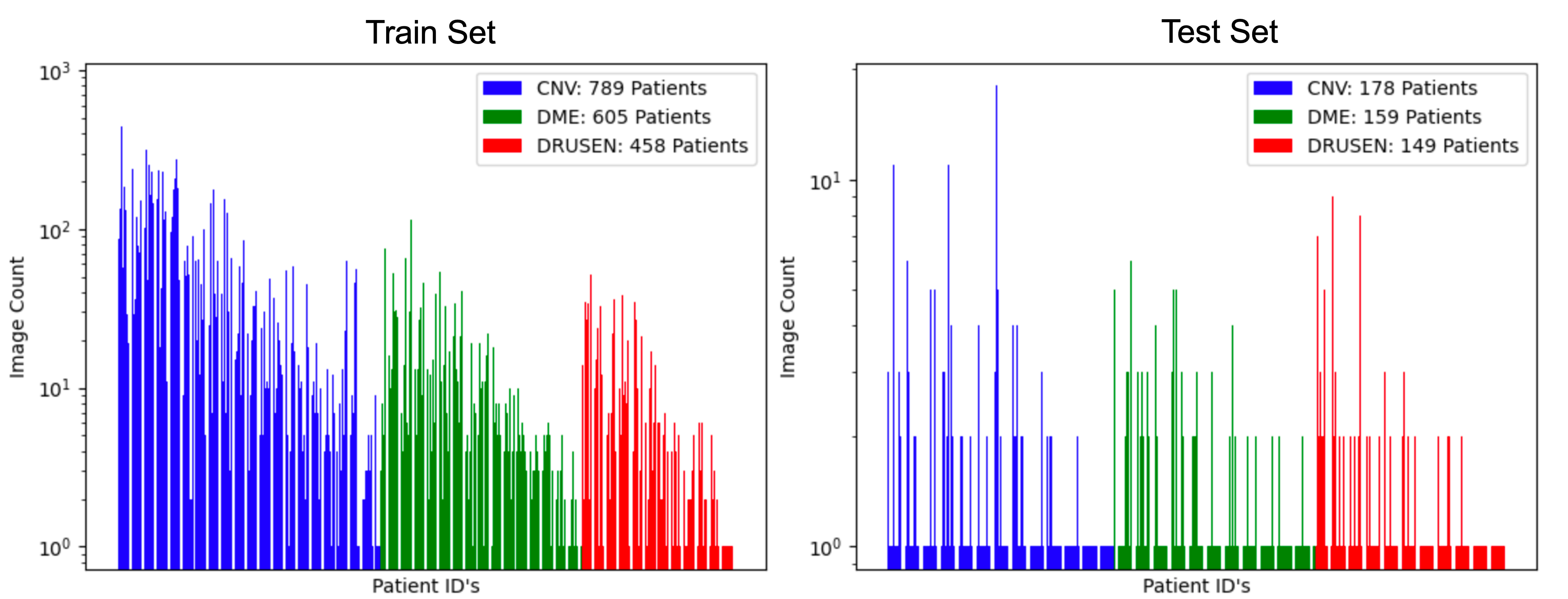}
\caption{Patient distribution for the OCT dataset.}
\label{fig:oct_dist}
\end{figure}

The dataset used in this paper, obtained from \cite{kermany2018large}, has grayscale, cross-sectional, foveal OCT scans of varying sizes belonging to an unbalanced distribution of a healthy class and three types of retinal diseases: Drusen, choroidal neovascularization (CNV) and diabetic macular edema (DME). We are interested in distinguishing between the disease states and thus use only imagery from these three classes for fine-grained classification. Sample imagery from each class is shown in Fig. \ref{fig:oct_pics}. A total of 10488 DME, 36345 CNV and 7756 Drusen images from 1852 patients were used in the training and unlabeled set. Within the test set, there were 250 images for each disease collected from 486 patients. All imagery were resized to $128 \times 128$ and normalized to have zero mean and unit standard deviation. Additional implementation details are shown in Table \ref{tab:implementation details}. There was no overlap in patients or imagery in train or test sets. The patient and class distribution in train and test sets are shown in Fig. \ref{fig:oct_dist}. 

\begin{table}[h!]   
  \begin{center}
    \caption{Details about dataset size and the number of samples added after each training round.}
    \label{tab:implementation details}
    \begin{tabular}{*{13}{c|c}}
      \toprule 
      \multirow{2}{*}{\textbf{Details}} & \textbf{Dataset}\\
       & OCT  \\
      \hline
      \midrule 
      Unlabeled + Training Set Images  & 54589  \\
      Images in Initial Training Set & 128  \\
      Images Queried per Iteration & 128 \\
      \bottomrule 
    \end{tabular}
  \end{center}
\end{table}

We use Resnet-18 \cite{he2016deep} and Densenet-121 \cite{huang2017densely} architectures trained from scratch to classify images in an active learning framework. A learning rate of 1.5e-4 was used along with the Adam \cite{kingma2014adam} optimizer. During each training round, the Resnet and Densenet models were trained for as many epochs it took to achieve 98\% and 94\% accuracy respectively. This is repeated with five different random seeds and an average accuracy is aggregated.

Classification performance is evaluated on the following baseline algorithms:
\begin{enumerate}[leftmargin=0.8cm]
    \item \textit{Random}: Randomly sampling $k$ instances at each round as a naive baseline.
    \item \textit{Least Confidence}: Selecting the samples for which the model has the lowest predicted probability \cite{settles2009active}.
    \begin{align}
    x^*_{LC} = \argmax_x 1 - P_{\theta}(\hat{y}|x)
    \label{eq:least_conf}
    \end{align}
    \item \textit{Margin}: The difference between the top two most probable predictions, $\hat{y_1}$ and $\hat{y_2}$, is used to identify the most informative sample \cite{settles2009active}.
    \begin{align}
    x^*_{M} = \argmin_x P_{\theta}(\hat{y_1}|x) - P_{\theta}(\hat{y_2}|x)
    \label{eq:least_conf}
    \end{align}
    \item \textit{Entropy}: Uses the full distribution to determine the most uncertain prediction from a model \cite{settles2009active}.
    \begin{align}
    x^*_{E} = \argmax_x \sum_i P(y_i|x)\log{P(y_i|x)}
    \label{eq:least_conf}
    \end{align}
    \item \textit{BADGE}: Samples disparate and high magnitude points from a gradient space at every round \cite{ash2019deep}.
\end{enumerate}

%% file: results.tex
\begin{figure} [h!]
     \centering
     \begin{subfigure}[b]{0.49\columnwidth}
         \centering
         \includegraphics[width=\columnwidth]{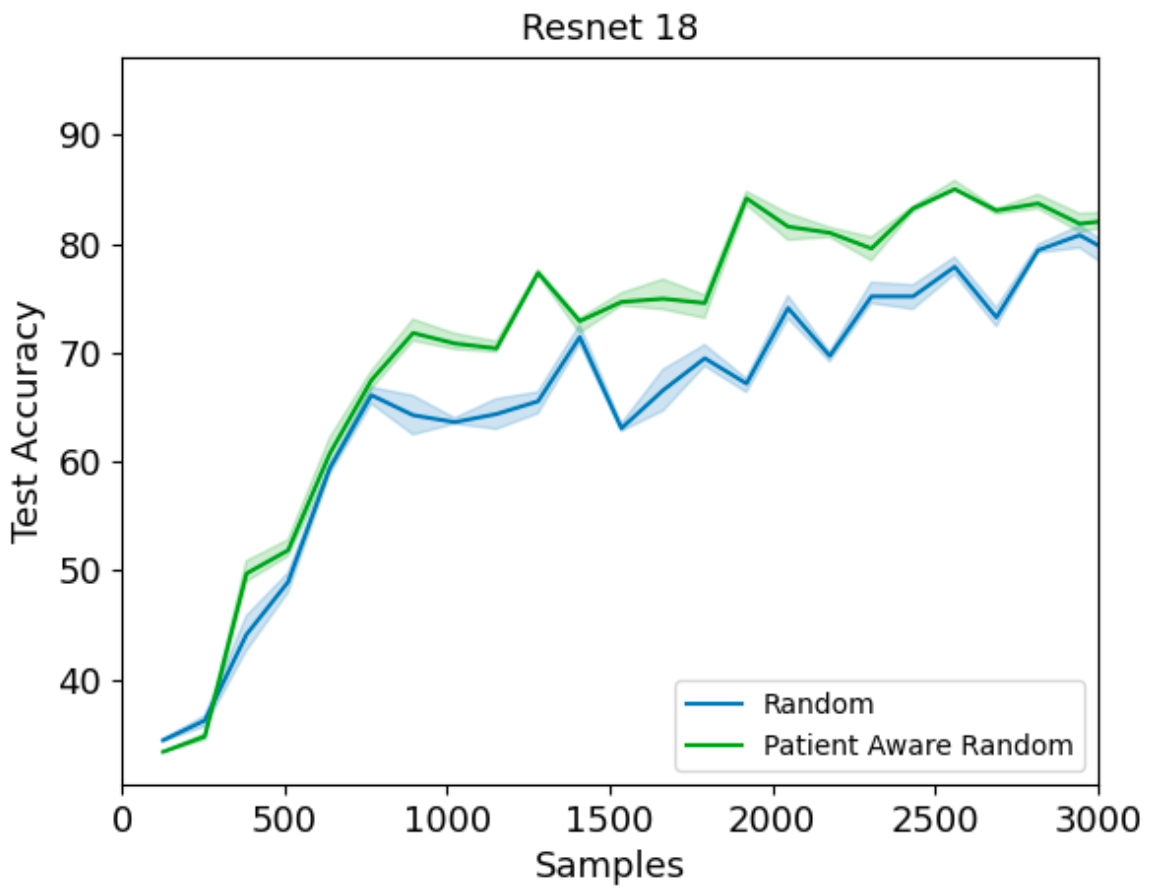}
         \caption{}
         \label{fig: oct rand}
     \end{subfigure}
     \begin{subfigure}[b]{0.49\columnwidth}
         \centering
         \includegraphics[width=\columnwidth]{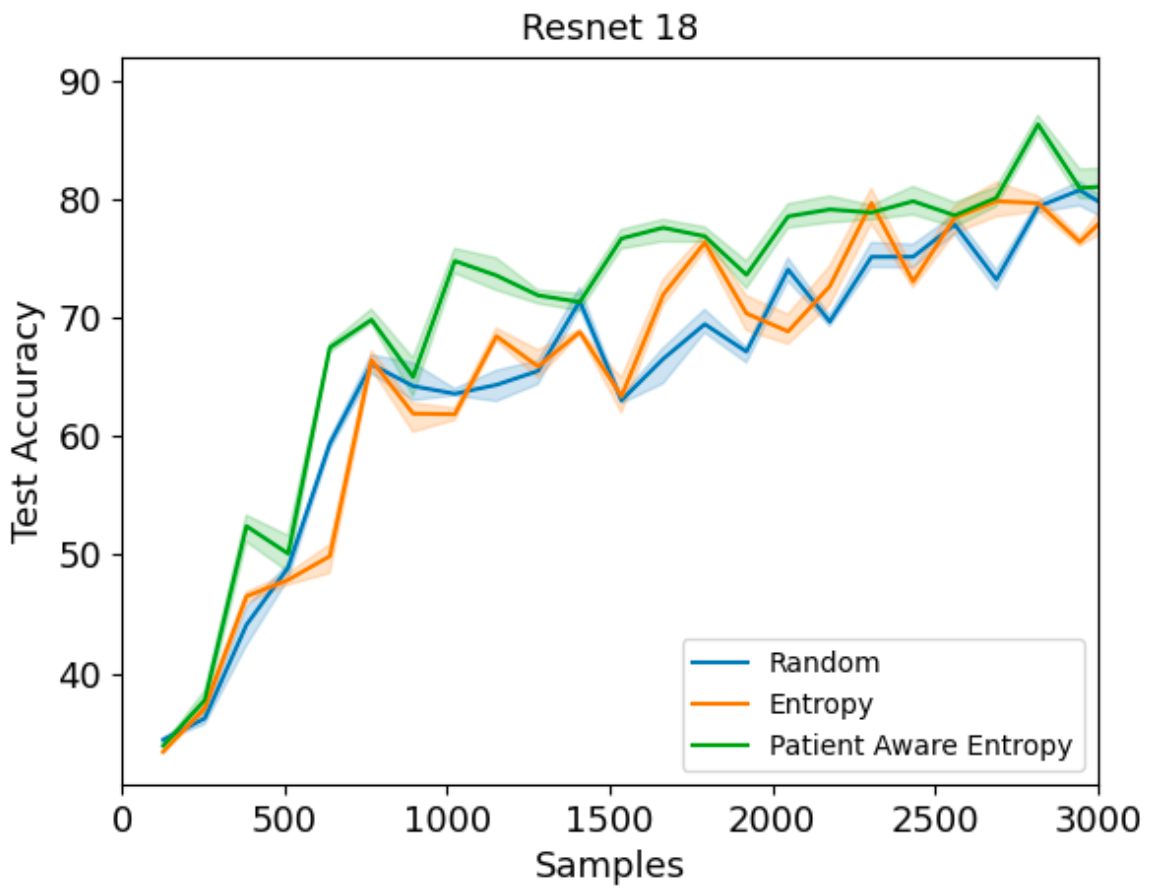}
         \caption{}
         \label{fig:oct entropy}
     \end{subfigure}
     \hfill
     \begin{subfigure}[b]{0.49\columnwidth}
         \centering
         \includegraphics[width=\columnwidth]{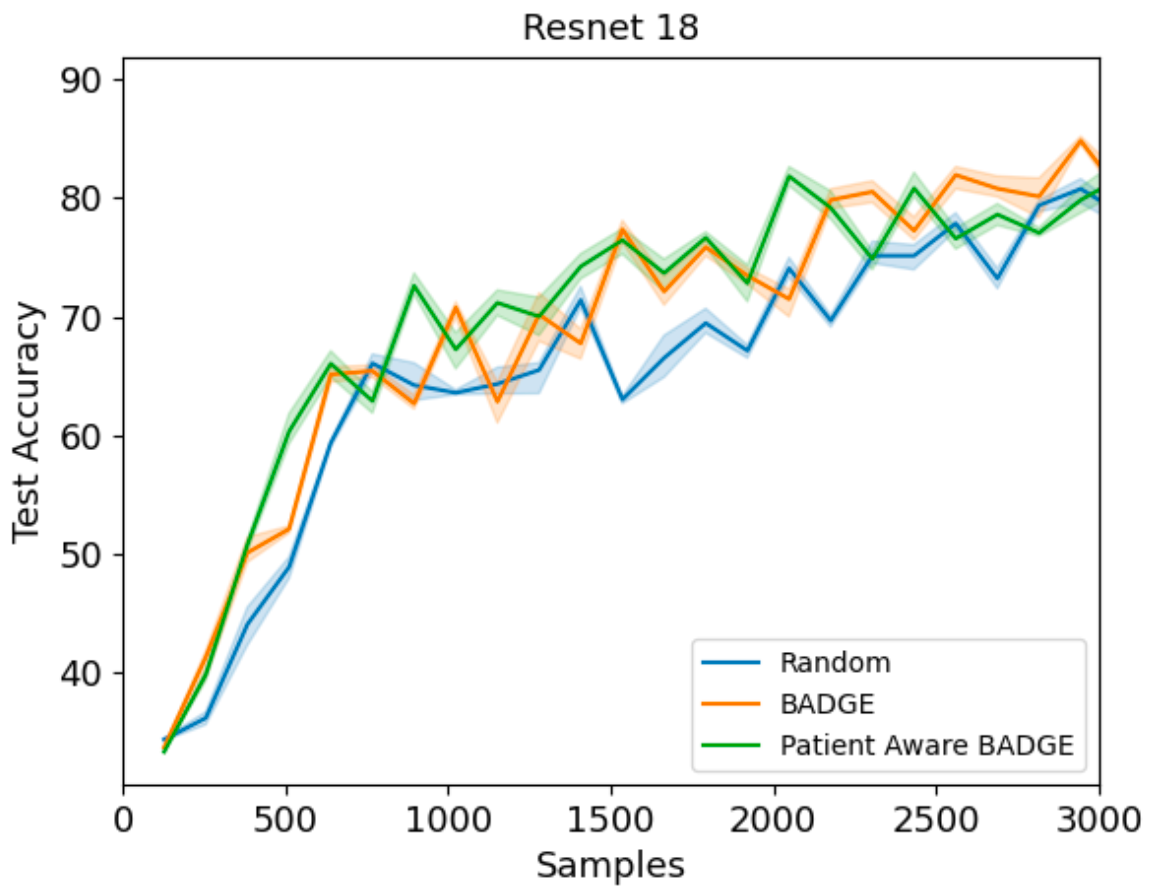}
         \caption{}
         \label{fig:five over x}
     \end{subfigure}
     \hfill
     \begin{subfigure}[b]{0.49\columnwidth}
         \centering
         \includegraphics[width=\columnwidth]{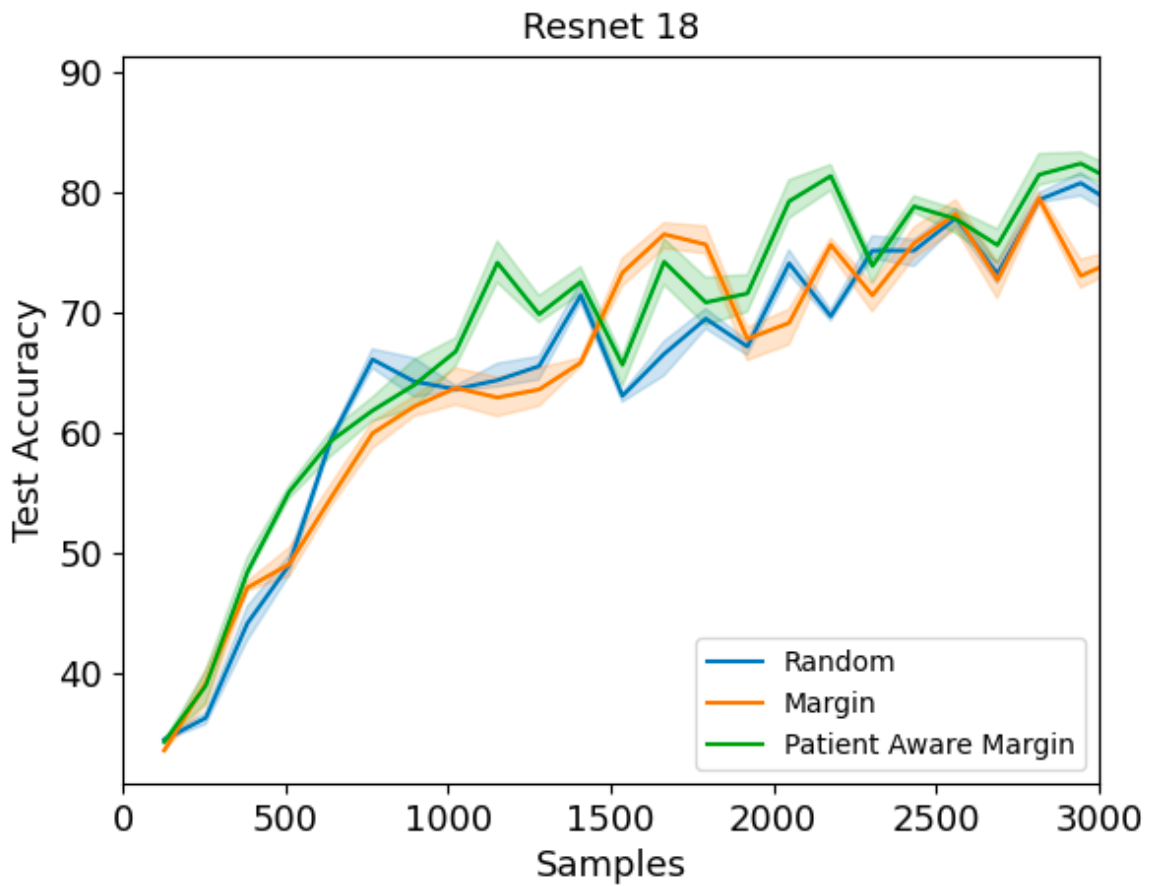}
         \caption{}
         \label{fig:five over x}
     \end{subfigure}
        \caption{Active learning test accuracy versus sample count for the OCT dataset on Resnet-18.}
        \label{fig:OCT curves resnet}        
\end{figure}
\begin{figure} [h!]
     \centering
     \begin{subfigure}[b]{0.49\columnwidth}
         \centering
         \includegraphics[width=\columnwidth]{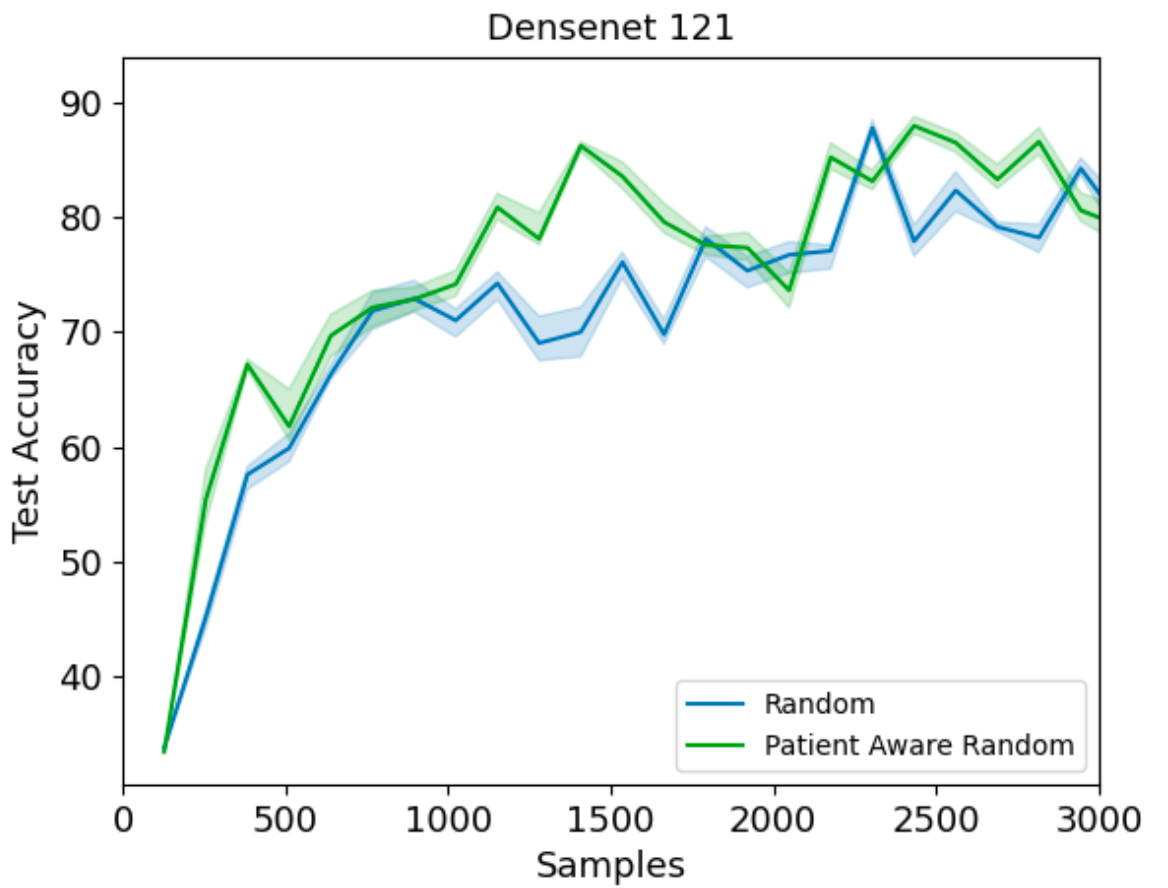}
         \caption{}
         \label{fig: oct rand}
     \end{subfigure}
     \begin{subfigure}[b]{0.49\columnwidth}
         \centering
         \includegraphics[width=\columnwidth]{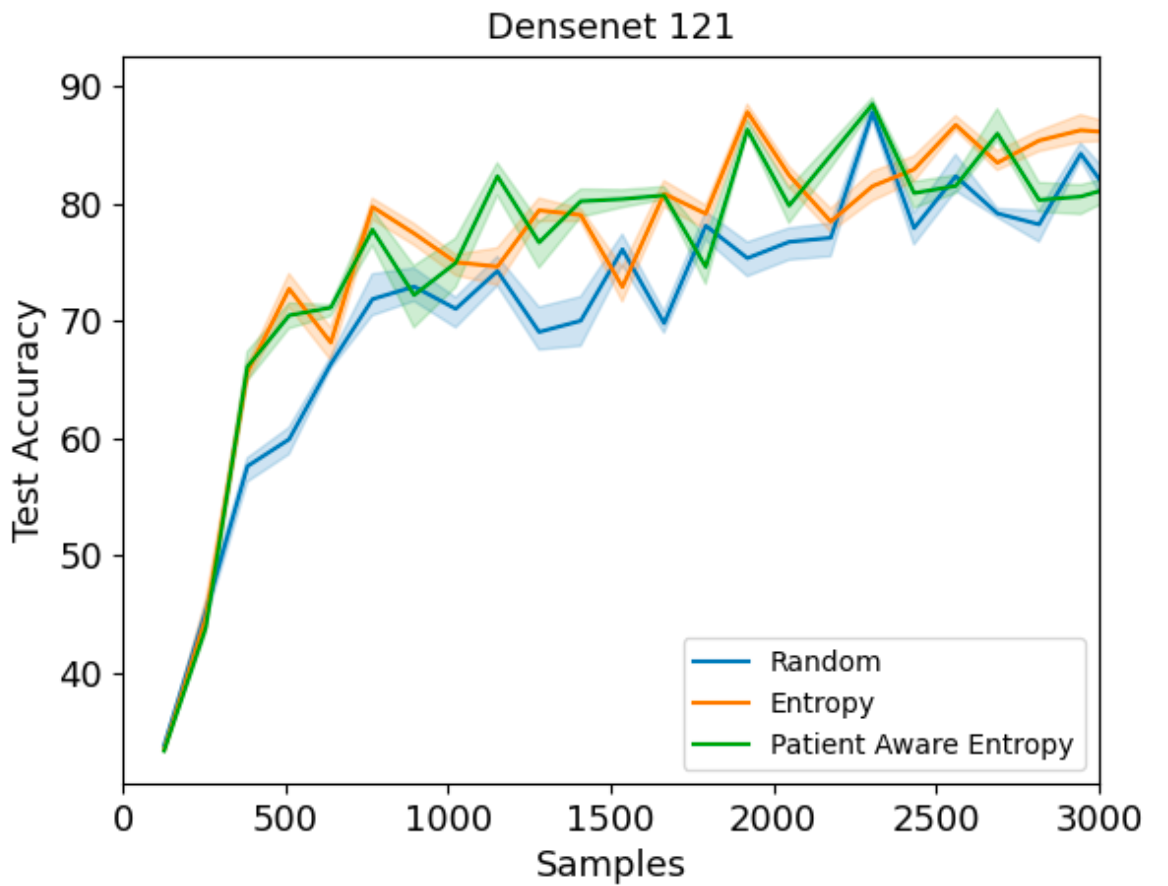}
         \caption{}
         \label{fig:oct entropy d}
     \end{subfigure}
     \hfill
     \begin{subfigure}[b]{0.49\columnwidth}
         \centering
         \includegraphics[width=\columnwidth]{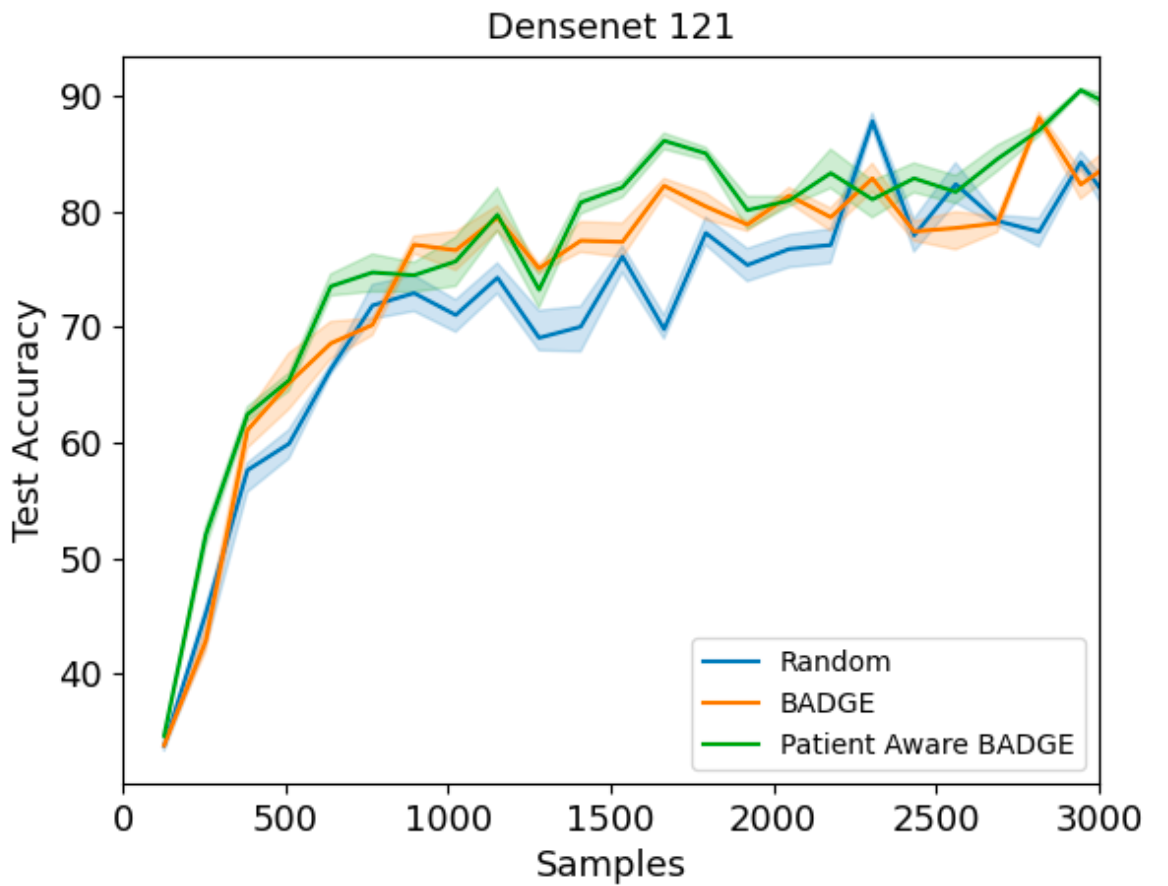}
         \caption{}
         \label{fig:oct badge d}
     \end{subfigure}
     \hfill
     \begin{subfigure}[b]{0.49\columnwidth}
         \centering
         \includegraphics[width=\columnwidth]{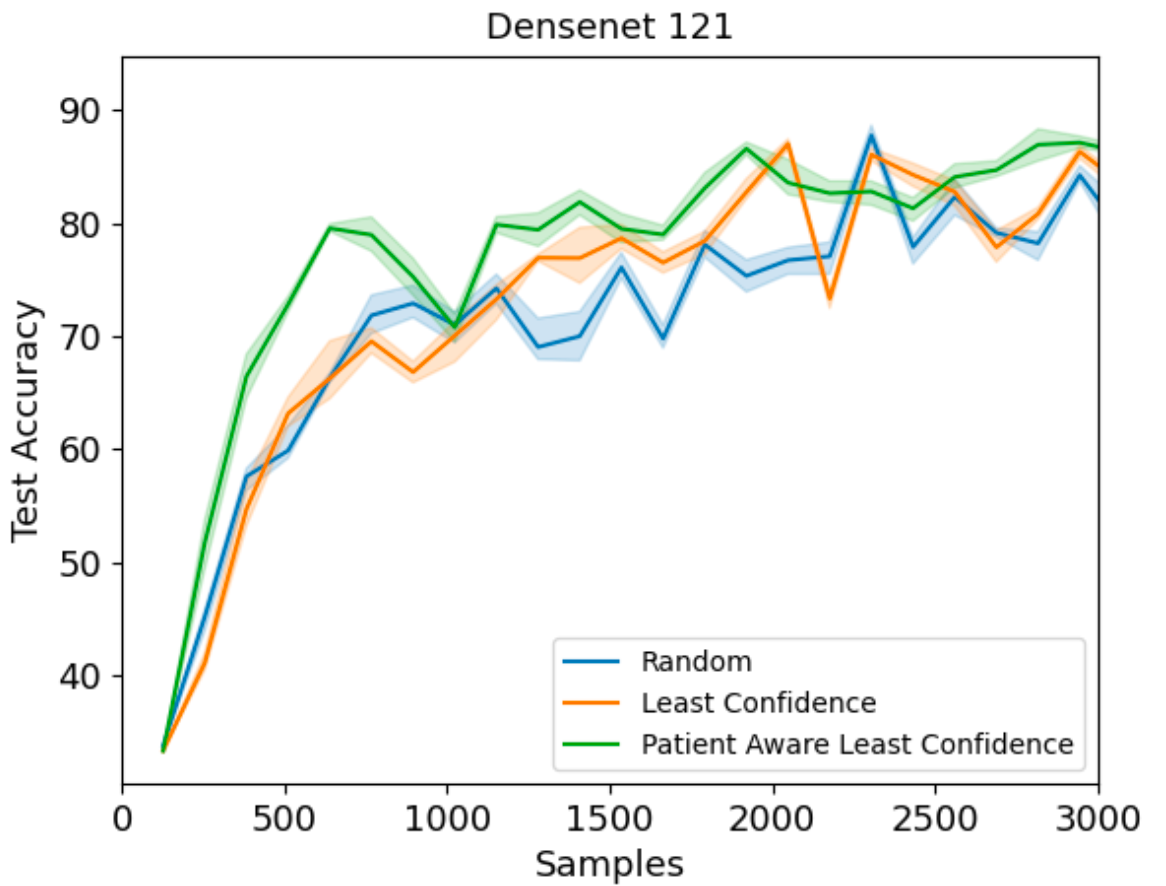}
         \caption{}
         \label{fig: oct lconf d}
     \end{subfigure}
        \caption{Active learning test accuracy versus sample count for the OCT dataset on Densenet-121.}
        \label{fig:OCT curves densenet}        
\end{figure}

In Figs. \ref{fig:OCT curves resnet} and \ref{fig:OCT curves densenet} we show learning curves showing the accuracy of baseline query strategies compared to the corresponding patient aware query. In the interest of space, we visualize results for four of five query strategies on Resnet-18 and Densenet-121 architectures. In these plots, x-axis corresponds to the number of samples in the train set at that round and y-axis corresponds to the performance accuracy on the test set. Each colored curve is the average of five trials using different seeds, with standard errors being shown by the shaded regions. Our patient aware learning applied to the baselines is shown as the green curve in all sub-figures while baseline queries are shown in orange. The blue curve represents random sampling which serves as the naive baseline for all experiments. All results were achieved using 5.5\% (3000 samples) of the OCT dataset (54589 samples).

The learning curves provide intuition about having medical insights guide the decisions of neural networks. For instance, most plots show patient aware learning surpassing the baselines on both architectures. This means that in majority cases, querying the dataset with patient-level insights in an active learning setup allows the model to better characterise disease states. Furthermore, we often see patient-aware learning having an edge over baseline algorithms from the early rounds of training onward like in Figs. \ref{fig:oct entropy} and \ref{fig: oct lconf d}. This is because the role of patient awareness plays a critical part in incorporating diversity throughout the training set. The model is thus exposed to several manifestations of a pathology from the onset. Noticeably too, patient aware learning can sometimes perform the same as baseline strategies, as shown in Fig. \ref{fig:oct entropy d}. This can happen when the selection of diverse patients do not contain significantly different disease manifestations within the imagery for the model to perform much different from the baselines. This is also possible when the model in some trials is poorly initialized yielding a poor set of built-in assumptions to make predictions. Overall, these findings suggest that patient aware learning is a good choice for medical applications of active learning.

%% file: conclusion.tex
In this paper we introduced a framework that incorporates clinical context in the form of patient awareness into an active learning framework that is medically interpretable. We show that our methods, in a plug-in sense, can be used in a modular context with arbitrary sampling strategies. We perform controlled experiments to validate the effectiveness of patient aware active learning for fine-gained OCT classification on different architectures using an unbalanced dataset. Based on our contributions in this paper, we will further investigate learning methods to evaluate the robustness of patient aware active learning.